# Feminism, gender identity and polarization in TikTok and Twitter

**Simón Peña-Fernández[1]\*, Ainara Larrondo-Ureta[2] and Jordi Morales-i-Gras[3]**

1. University of the Basque Country (UPV/EHU) (ROR: 000xsnr85)
   simon.pena@ehu.eus / https://orcid.org/0000-0003-2080-3241
2. University of the Basque Country (UPV/EHU) (ROR: 000xsnr85)
   ainara.larrondo@ehu.eus / https://orcid.org/0000-0003-3303-4330
3. University of the Basque Country (UPV/EHU) (ROR: 000xsnr85)
   jordi.morales@ehu.eus / https://orcid.org/0000-0003-4173-3609

\* Corresponding author

**Abstract**: The potential of social media to create open, collaborative and participatory spaces allows young women to engage and empower themselves in political and social activism. In this context, the objective of this research is to analyze the polarization in the debate at the intersection between the defense of feminism and transsexuality, preferably among the young population, symbolized in the use of the term "TERF". To do this, the existing communities on this subject on Twitter and TikTok have been analyzed with Social Network Analysis techniques, in addition to the presence of young people in them. The results indicate that the debates between both networks are not very cohesive, with a highly modularized structure that suggests isolation of each community. For this reason, it may be considered that the debate on sexual identity has resulted in a strong polarization of feminist activism in social media. Likewise, the positions of transinclusive feminism are very much in the majority among young people; this reinforces the idea of an ideological debate that can also be understood from a generational perspective. Finally, differential use between both social networks has been identified, where TikTok is a less partisan and more dialogue-based network than Twitter, which leads to discussions and participation in a more neutral tone.

## 1. Introduction

*1.1. Feminism on social media*

Social media has considerable potential to involve and empower young women in political and social activism (Batsleer & McMahon, 2016), thanks to its capacity to create open, collaborative and participatory spaces for feminism (Ott, 2018). Platforms such as Twitter and TikTok allow young women freely to express themselves and converse with all manner of social agents to express their opinions and feelings. They are also able to exchange information as to the issues involved in the construction of sexual identities, or gender-based injustices that they have experienced or witnessed (Jackson, 2018).

The spread of social media has likewise allowed the feminist movement to create and raise awareness of a host of issues, whether it be sexism, inequality, or gender violence (Baer, 2016), serving to extend the scope of the movement's claims. In this regard, Social Sciences drive to advance and address parameters which go beyond the influence which can be measured in terms of the number of followers and retweets, because in a hyperconnected and interactive world, social activism is, above all, a conversation. The trends encapsulated within what is known as the "affective turn" in this academic field (Ticineto-Clough & Halley, 2007) also plays an important role regarding this epistemological advance, as has the momentum given to certain activisms online.

Among those social movements that have developed on social media, feminism has revealed a prominent role, as highlighted by gender studies conducted in conjunction with sociology, pedagogy, and communication. Twitter, TikTok and other social media platforms have communicative characteristics (immediacy, media impact, message simplification, mobilization capacity, etc.) which favour the creation of ideologically like-minded communities. As demonstrated in the case of what is known as "feminist hashtivism", through such high-profile campaigns as #Metoo, #WomensMarch, etc. (Jinsook, 2017; Turley & Fisher, 2018; Storer & Rodriguez, 2020; Linabary et al., 2020).

According to the most recent Report on Youth in Spain (Injuve, 2021), the interest among the younger population regarding gender inequalities could be connected with the fact that they are socialising within a context in which the main mobilizations are linked to the feminist cause. In parallel, there has been, over recent years, an increase in the number of critical voices demanding an intersectional reading of online activism, as the very concepts of cyberfeminism and gender have been overtaken by pressing social changes and debates in the virtual world (Salido-Machado, 2017). In this sense, the feminist movement has succeeded in arousing the interest of a younger population, pushing them towards mobilization in the offline and online sphere, but in particular in the latter.

This type of analysis proves of interest in exploring in greater depth the mechanisms allowing young people, as a group, to construct and socialize their identity or feminist consciousness through all manner of systemic and counter-systemic positions via online media platforms. What is more, an analysis of such media proves of interest in order to understand positions or discourses of hate based on dialectic confrontations, which currently represent a sphere which has still not been explored to any great extent in the literature focused on expressions of the identity of the young population in connection with gender in digital environments.

*1.2. Polarization of discourse as to identity*

In the contemporary feminist movement, there has been an intense debate about identities and their political subject, characterized by the extensive use of digital platforms (Willem & Tortajada, 2021). The inclusion within the feminist struggle of claims linked to environmentalism and the LGTBI collective has deep roots and a lengthy theoretical discussion behind it (Earles, 2017). The debate about identities in feminism is a complex one, and trans issues and the very language used by different agents in these debates not only constitute disputes in terms of terminology or how sex and gender should be conceptualized, but are also forms of representing a positioning in this dispute. Trans/feminist conflicts, also known as "TERF wars", reflect the current conditions of our day, in which public discourse is dominated by polarization and the proliferation of disinformation.

What is known as Trans-Exclusionary Radical Feminism (or "TERF") is one of the most widespread terms within the context of the digital feminist debate. The term was first used around the year 2008 and is now very commonly employed in digital conversations on such social media platforms as Twitter (Sulbarán, 2020). The term "TERF" is there used with negative connotations linked to the alleged transphobia shown by feminists who identify with their biological gender, towards transgender women or women who do not self-identify with their biological agenda. In this regard, feminists classified as "TERF" perceive this term as having negative connotations, or even as an insult (Malatino, 2021). The academic literature includes recent studies which examine the rise of the "TERF" anti-transgender movement in English-speaking contexts (Mclean, 2021).

Recent works such as "TERF wars: An introduction" (Pearce et al., 2020) have highlighted the intensity of these debates or "dialectic wars" on social media, as well as their importance in extending our understanding of the trans phenomenon from the perspective of a younger population, through the analytical framework of cyberfeminism, and also from the ideological context from which trans-exclusionary arguments emerge.

The discourses regarding trans-inclusive positions have been highly polarized (Carrera-Fernández & DePalma, 2020), generating an interest in understanding the way in which these debates spread via social media. The dialectical struggle between the different concepts as to the political subject of feminism and the location of identities within this sphere have been investigated on such social media platforms as Instagram (Vázquez-González & Cárdenes-Hernández, 2021), Twitter (Lu, 2020) and YouTube (Tortajada et al., 2021).

The analytical and conceptual framework which has thus far been provided by studies linked to digital feminism is becoming increasingly extensive. Among these analyses, we here highlight the case investigation into the hashtag #ContraElBorradoDeLasMujeres ("#AgaintTheDeletionOfWomen"), indicating that the discourse generated in connection with this hashtag is highly emotive (Ferré-Pavia & Zaldívar, 2022). Similar studies highlight that this type of tag reinforces positions and serves to identify discourses in the digital public debate, in particular among a younger population, as the main user group of the spaces covering this type of dialogue. As set forth in various studies, social media users tend to become majorities with radical positions thanks to the influence of phenomena of reciprocal influence, such as polarization and echo chambers (Demszky et al., 2019). The existence of parameters or indicators such as language proves useful in these analyses to detect affinities in the conversation or, where applicable, polarities determining the democratic and social usefulness of public digital spaces.

Regarding the topic which here concerns us, "feminist hashtivism" has been upheld in general terms as an enhanced and enriched digital feminist activism, drawing on theoretical-conceptual artefacts such as intersection, colonialism, multimodal violence, etc. This phenomenon began to emerge prominently as a result of the #MeToo mobilizations and represented a cornerstone not only for other digital feminist mobilizations in combating all forms of sexist discrimination (violence, pay inequality, discrimination of any type, etc.), but also for all manner of research into the value of social media for this movement (Manikonda et al., 2018).

Feminism is of interest in this regard, since it defines itself as a movement with a theoretical corpus which is not closed, and with a practice which evolves towards new forms of action and protest, such as those occurring in different realms and contexts of the digital public-political sphere. It is no coincidence that young people typically encounter hatred in their activities online, and debates concerning LGBTI issues are among the most affected (Council of Europe, 2014; Injuve, 2019).

In this context, the objective of this research focuses on an analysis of activism and polarization arising in the social media debate at the intersection between the defense of feminism and transsexuality, above all among a younger population, symbolized in the use of the term "TERF" (Trans-Exclusionary Radical Feminism). On this basis, the following research questions are raised:

RQ1. Which social media communities have been created in connection with feminist activism in terms of the inclusion of demands connected with gender identity?
RQ2. What is the presence of young people in these communities, and their position regarding this matter?
RQ3. What is the degree of polarization in the debate within feminist activism on social media concerning this topic?

## 2. Methodology

In order to measure the polarization which occurs in feminist activism on social media, in particular among young people, an analysis was conducted of the use of the term "TERF" on TikTok and Twitter between 5 March and 11 March. In other words, data were compiled for 8 March ±3 days, taking advantage of the increased debate occurring on social media regarding topics connected with feminism as a result of International Women's Day.

The choice of Twitter follows on from prior research into polarization and feminism on this platform, and the influence that it acquires as the soapbox used by public figures to express their opinions. Twitter has 4.2 million registered accounts in Spain, of which 796,385 are considered active accounts (having published content in the last two months) (Social Media Family, 2022). By user gender, men (32%) slightly outnumber women (28%), although 40% do not specify their gender. Meanwhile, the choice of TikTok is based on the fact that this is a social media platform with a very young audience, 41.4% of users being aged between 18 and 24, while 59.3% are women. On this platform, the most popular category is News and Entertainment.

Twitter data were accessed by means of the API 2.0 with academic access, which serves to analyze a large volume of data defined both in semantic and temporal terms. Meanwhile, the TikTok data were acquired by means of web scraping techniques, as the platform does not yet provide access to an academic API, at least for the moment (TikTok, 2022). The difference in the conditions of access to data on Twitter and TikTok is the first limitation of the study which must be taken into account in terms of methodological design itself. Once the data were downloaded, both conversations were then analyzed by means of Social Media Analysis techniques with a twofold objective: 1) identify structural dynamics on both platforms to ascertain the uses made of digital content by defined user groups; and 2) facilitate basic conditions to compare what happens on the two platforms, for which the same data access conditions are not available.

It should lastly be emphasized that the term "TERF" itself is controversial and is not used in this research in its descriptive or characterizing facet, but as a core around which debates concerning sexual identity can be identified on social media, in the context of digital feminist activism. In this regard, the second limitation of this study lies in this differential identification with the term and the intentionality with which it is used, which influences the frequency of its use by each of the communities, and which does not cover the social media debate as to sexual identity and its relationship with feminism to its full extent.

*2.1. Data analysis on Twitter*

For the Twitter analysis, 24,714 tweets with the term "TERF" in the singular and plural were captured during the period analyzed. The conversation was subsequently converted into a directed network of retweets in which each node represents a user who retweeted or was retweeted by another. In total, the network comprised 10,449 nodes which retweeted with one another 10,970 times. Of these, 8,666 nodes did not receive any retweet, and 1,478 did not retweet to anyone else. Only 305 nodes (2.92% of the total) sent and received at least one retweet. If we take into account only those nodes that retweeted to another node, thus discarding those that only received retweets, the average number of retweets per user is 1.27.

The first analytical step involved identifying the user groups that conducted conversations among themselves. Following application of the Louvain algorithm (Blondel et al., 2008) for community detection (RQ1) with NetworkX (Hagberg et al., 2008) for Python, 980 communities were identified (Figure 1).

To estimate age (RQ2), the baseline situation is that the Twitter API does not provide data as to users' ages. Any algorithm-based inference is essentially precarious, since it requires declarative data from users who do not always exist, or advanced data-mining procedures which, although they may ultimately be precise, are always very difficult to corroborate.

Lastly, to estimate the average age of the nodes of each cluster, a different strategy was employed, based on the artificial intelligence developed by Twitter itself to classify its users. The creation of two advertising campaigns was thus simulated, respectively segmenting users similar to the trans-inclusive and non-trans-inclusive cluster nodes, comparing the estimations of their scope by age band. Segmentation of users sharing the node characteristics and differentiated by age band thus served to estimate the support for each of the two positions among young people. The age bands that Twitter allows for segmentation are 13 to 24 years, 21 to 34, 35 to 49, and over 50.

This mode of segmentation has its limitations. To begin with, the age bands used by Twitter partially overlap, since two of them include young people aged between 21 and 24. This is because the tool is not intended for scientific research, but rather for audience segmentation in advertising campaigns. Lastly, the discourse of each of the communities was characterized (RQ3) on the basis of the content analysis of the most retweeted messages, and of the most followed accounts and the tags most shared by these communities.

*2.2. Data analysis on TikTok*

To analyse the TikTok data, the strategy deployed was different from the case of Twitter because of the differences between the two platforms in the type of communication they permit between users, and in the access to data. While in the latter the interactions are based on likes, retweets, replies, quotes and other types of reference (with a retweet being the interaction best denoting affinity between users), on TikTok the interactions essentially comprise likes and replies to content uploaded by users in video form. While on Twitter the information was obtained via the academic

API, which offers very good data access, even retroactively, on TikTok, data can only be obtained by means of web scraping techniques.

To conduct the analysis, data were downloaded for the 500 most viewed videos containing the term "TERF" with PhantomBuster. Of these, 165 (33%) contained descriptions in Spanish, the comments on which were downloaded with the Google Chrome Web Scraper plug-in. This technique provided access to the comments directly posted to the video, setting aside replies to the comments. In total, 16,974 comments were downloaded, and analyzed by means of Social Media Analysis techniques, with each network node being a user, and each link corresponding to one or more comments made by one user towards another. Lastly, the network analyzed comprised 12,687 nodes and 16,091 edges.

## 3. Results

*3.1. Communities and polarization*

In the case of Twitter, the 24,714 tweets analyzed served to identify 980 communities, with a Modularity of 0.939. This parameter measures the quantity of fragmentation of a graph, presenting values close to 0 when the graph comprises one single community in which all nodes are symmetrically and horizontally related, and values close to one when the nodes tend to comprise fragmented communities without contacts among them. These figures thus show that the conversation is highly fragmented and that each community is individually isolated. For this topic, then, Twitter proves to be a network with very little cohesion, and user opinions are most likely polarized (Figure 1).

Figure 1. Network of retweets (Twitter) and comments (TikTok)

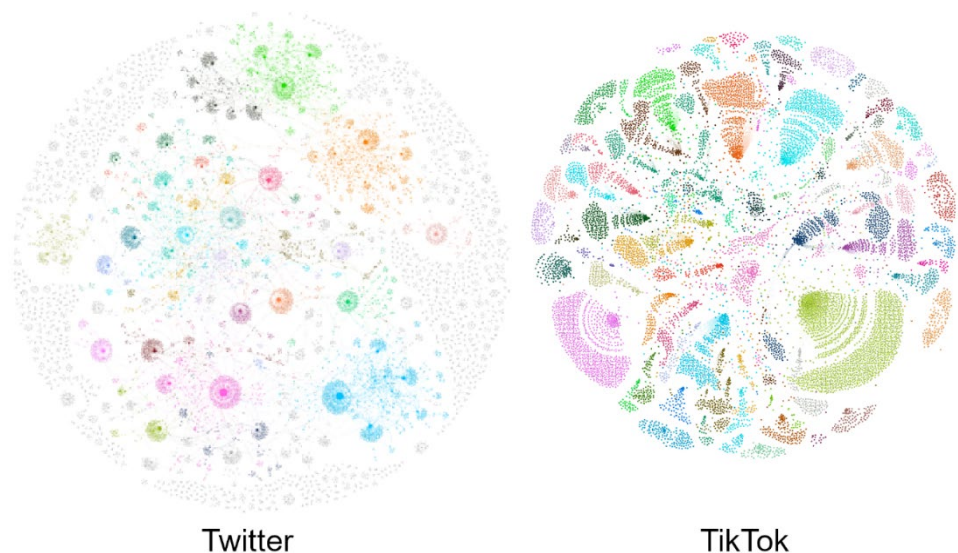

Source: Author's own work.

After applying the same Louvain algorithm to the network of comments on TikTok, we note that the network of replies shares some structural characteristics with the network of retweets on

Twitter, such as a high degree of modularity, of 0.854 points, and an architecture revealing little cohesion, with each community being individually isolated. This means that on both social media platforms, interactions in the "TERF" debate tend to take place within closed circles. The network density was 0.000097 on Twitter and 0.0001 on TikTok. This means that on both platforms, the vast majority of possible links between nodes remain unexplored: users prefer to interact with very small groups of other users, thereby limiting their form of participation, and eschewing more broad-based debates.

Figure 2. Nodes by position

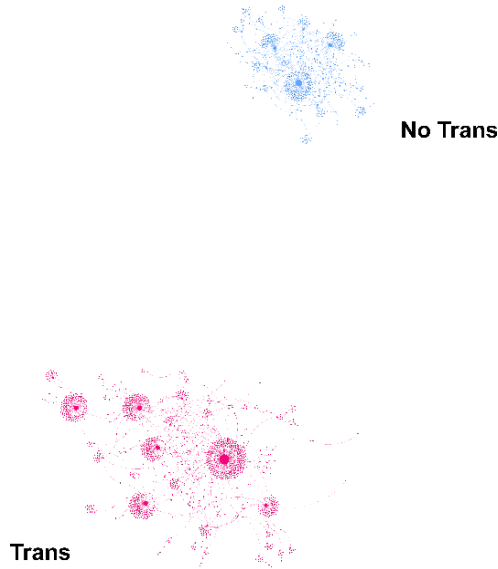

Source: Author's own work.

To identify which modes of cohesion operate within the network, a series of categories were synthesized, classifying nodes in accordance with four variables: the language most used by each node in its retweets, the language most used in each community, the provenance of the users (Spain or elsewhere), and within the Spanish communities, whether the position upheld is or is not inclusive. All these categories were visualized (Figure 2) and evaluated by means of the categorical assortativity statistic available in NetworkX (Hagberg et al., 2008), this being one of the most typical forms to measure intra-community homophily in social media political polarization studies (Leifeld, 2018; Taylor et al., 2018; Salloum, 2021) (Table 1).

Table 1. Categorical assortativity by different categories on Twitter

| Category | Categorical assortativity |
| --- | --- |
| Language of the user | 0.813 |
| Main language of the community | 0.991 |
| Main country in the community (Spain vs. Others) | 0.969 |
| Position, for the Spanish communities (Inclusive/Non-inclusive) | 1.0 |

Source: Author's own work.

All the categories considered proved to be highly assortative. Evidence is thus found of linguistic, national and ideological homophily in a highly polarized network in all assortativity indicators consulted. Language, whether measured individually for each node or measured according to the majority use of each community, proves a fundamental element in explaining the relationship between nodes. Users tend to retweet content always in the same language, and communities likewise tend to be formed in accordance with criteria of linguistic homogeneity.

We can likewise see that users belonging to mainly Spanish communities tend to relate to like communities, and to isolate themselves from different communities. Within the Spanish communities, total assortativity is also found with regard to the two positions: the nodes with trans-inclusive and non-trans-inclusive feminist positions do not relate to one another.

Although the statistics to evaluate polarization can only be applied to Twitter, the data show that both networks have little cohesion, since the nodes are mainly disconnected from one another, and connected only to the most retweeted users or to the content creators. They furthermore have a highly modularized structure, with a modularity figure close to 1, which suggests that each community is individually isolated.

*3.2. Young women*

Positioning by age was established directly on the basis of simulation conducted in Twitter by means of its marketing tool, and indirectly by means of the user profiles on each of the social media platforms analyzed.

In the former case, Twitter shows that trans-inclusive feminist positions are very much the majority among young women, as we may estimate according to the data analyzed that three in every four users supporting this option are aged under 34 (Figure 3). Meanwhile, 45.4% of feminists upholding non-trans-inclusive positions are aged 35 or over, and in the segment aged over 50, non-inclusive positions outnumber the opposite view almost 5 times over. In the light of these data, it may be asserted that the supporters of this discourse on social media have an older audience, reinforcing the idea of an ideological debate at the heart of the feminist movement, which may also be understood in generational terms.

Figure 3. Distribution by age according to positioning

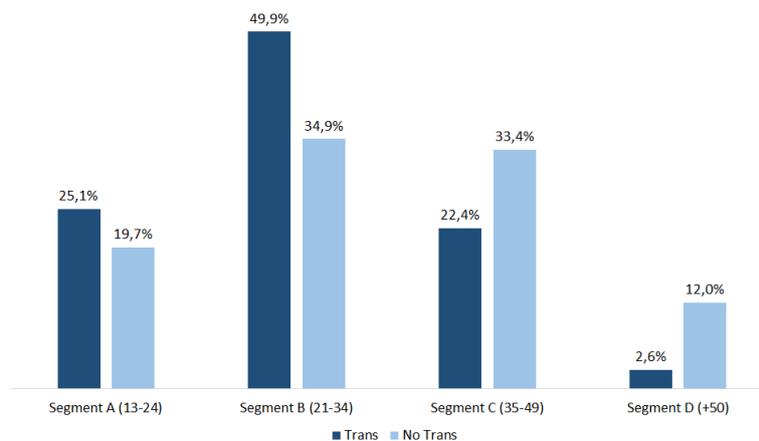

Source: Author's own work.

*3.3. Discourse*

On Twitter, the community which refers to itself as "antiTERF" articulates a confrontational, conflictive and even violent discourse in terms of the symbology deployed. There are constant suggestions that non-trans-inclusive feminists should be isolated and expelled from International Women's Day marches, even denying that there is a place for them in the feminist movement. The non-inclusive discourse is presented in these communities as discriminatory, a breach of human rights, regarding which no form of debate is possible.

Table 2. Characterization of the most relevant networks on Twitter

| Community | Discourse | Nodes | % Nodes | Most flw. profiles | Most popular hashtags |
|---|---|---|---|---|---|
| 3 | Not Trans | 524 | 4.99% | @ContraBorrado @AranguezTasia @jk_rowling @Barbijaputa @peliradfem | #movimientofeminista #stopdeliriotrans #leytransesmisoginia #leytransesmaltratoinfantil #rompeelgeneronosucuerpo |
| 4 | Trans | 523 | 4.98% | @__erosgarcia @PutoMikel @firecrackerx @SaraRiveiro @PuteadoMikel | #eurovision #maestrosdelacostura #laisladelastentaciones #cyberpunkedgerunners #edgerunners |
| 11 | Trans | 199 | 1.90% | @IbaiLlanos, @badbixsamantha, @PutoMikel, @auronplay, @MisterJagger_ | #70ssiff #deltarune #steddyhands #theowlhouse #digitalart |
| 17 | Trans | 169 | 1.61% | @iamthekillerq, @badbixsamantha, @DragRaceEs, @hugaceo_cruji, @ladygaga | #eurovision #shehulk #cáncer #foreverlove #mewsuppasit |
| 20 | Trans | 149 | 1.42% | @IbaiLlanos, @natalialacunza, @_albxreche, @auronplay, @Aitanax | #laisladelastentaciones #eurovision #eurobasket #jaehyun #wordle |
| 21 | Trans | 147 | 1.40% | @badbixsamantha, @IbaiLlanos, @IreneMontero, @senorcito, @VelvetMolotov | #laisladelastentaciones #splatoon3mm #pictdle #elalfilyladama #discapacidad |

Source: Author's own work.

The discourse of the different communities that have been identified can be characterized on the basis of the most followed profiles and the most used tags. For example, in community 3 (Table 2), which stands out for its non-trans-inclusive discourse, the leading accounts include @ContraBorrado, that of the influencer @Barbijaputa, and the writer @jk_rowling. The presence of this type of account, and the importance of hashtags opposing the Spanish legislation known as the "Trans Bill", bear witness to the high degree of activism within the community: this is a community made up of activists, in which other topics scarcely have a presence.

In communities aligned with trans-inclusive feminism, meanwhile, there are notable opinion-leaders associated with queer discourse, but one also finds leading figures among popular youth culture in Spain, such as the YouTubers Ibai Llanos, AuronPlay and Mister Jägger, who are not particularly connected with feminist activism on social media, beyond occasional statements. Both the type of leadership identified in these communities and the hashtags used (such as

#eurovision, #laisladelastentaciones, #eurobasket and #wordle) reveal a less activist use of Twitter on the part of the accounts of these communities, which include the topic in a more horizontal manner.

Table 3. Characterization of the most relevant networks on TikTok

| Community | Discourse | Nodes | % Nodes | Most commented profiles |
|---|---|---|---|---|
| 1 | Trans | 1785 | 14.07% | @hdeharva |
| 2 | Trans | 907 | 7.15% | @le_dudette |
| 3 | Not Trans | 614 | 4.84% | @joanne_fem |
| 4 | Trans | 593 | 4.67% | @_pic0tres, @shinji.anti.terfs |
| 5 | Not Trans | 471 | 3.71% | @soff.duh |
| 6 | Trans | 427 | 3.37% | @camradhoe, @bellafiera, @noagrcia_ |
| 7 | Trans | 369 | 2.91% | @gato.de.biblioteca, @rocioesperilla |
| 8 | Trans | 368 | 2.90% | @ilinkaandarcia |
| 9 | Trans | 368 | 2.90% | @melii_jade |
| 10 | Trans | 316 | 2.49% | @emmapalmina |
| 11 | Mixed | 296 | 2.33% | @lamejorcapricorniana, @ddaimode, @myqueerdom |
| 12 | Trans | 292 | 2.30% | @rebelarme |
| 13 | Not Trans | 290 | 2.29% | @lilith_131 |
| 14 | Trans | 289 | 2.28% | @gg_well_play |
| 15 | Trans | 241 | 1.90% | @maria_byw_love |
| 16 | Mixed | 208 | 1.64% | @sung.jinnie, @hell.alexaa |
| 17 | Trans | 206 | 1.62% | @aannaabbeeell, @cristinapadrolrov |
| 18 | Trans | 187 | 1.47% | @bravo_joack, @juancamiloreyes |
| 19 | Trans | 170 | 1.34% | @franchesqui__lopez |
| 20 | Trans | 158 | 1.25% | @luisii.4370 |
| 21 | Ambiguous | 156 | 1.23% | @mateotrosko |
| 22 | Trans | 152 | 1.20% | @diegocond1 |
| 23 | Ambiguous | 152 | 1.20% | @chocottete |
| 24 | Trans | 150 | 1.18% | @yosoymariamaya |
| 25 | Trans | 144 | 1.14% | @heyitschiquiar |

Source: Author's own work.

On Twitter, some of the messages most shared by trans-inclusive feminists characterize non-trans-inclusive feminists, whom they refer to as "TERF", as privileged, white, well-to-do and almost always heterosexual women. Although age does not form part of the set of privileges of non-trans-inclusive feminism, the presence of references to age in some of the messages serves to indicate that the generational element hovers over the dispute, with trans-inclusive feminists seeing themselves as the future of feminism.

On Twitter, meanwhile, the typical content of non-trans-inclusive discourse is to complain of the violence which they suffer at the hands of transactivists. It should be recalled that the concept of "TERF" itself, although it was created as an explanatory rather than derogatory term (Smyte, 2018) is considered an insult by most of if not all of the women accused of belonging to this category.

Whatever the case, when the term "TERF" is used in the communities we have here designated as non-trans-inclusive, this is to criticize the caricature which they see as being applied to them from trans positions. The argument is that the climate at marches is violent, that their activists are insulted and attacked, and that there is no possible reconciliation between the two positions. They even accuse their adversaries as serving as the weapon of age-old patriarchal violence against women.

In the case of TikTok, given the characteristics of the data comprising the network, the same cohesion and homophily tests cannot be performed as for Twitter, nor can we estimate either the geographical location or age of the users by means of the advertiser platform. Nonetheless, the comparative analyses that can be performed serve to reveal certain trends on this platform.

Table 4. Languages of conversation on Twitter

| Community | Nodes | % Nodes | Main language |
|---|---|---|---|
| 1 | 811 | 7.72% | Spanish |
| 2 | 572 | 5.45% | English |
| 3 | 524 | 4.99% | Spanish |
| 4 | 523 | 4.98% | Spanish |
| 5 | 309 | 2.94% | Spanish |
| 6 | 305 | 2.91% | English |
| 7 | 235 | 2.24% | English |
| 8 | 211 | 2.01% | Japanese |
| 9 | 207 | 1.97% | English |
| 10 | 201 | 1.91% | English |
| 11 | 199 | 1.90% | Spanish |
| 12 | 197 | 1.88% | Spanish |
| 13 | 191 | 1.82% | English |
| 14 | 190 | 1.81% | English |
| 15 | 178 | 1.70% | English |
| 16 | 173 | 1.65% | French |
| 17 | 169 | 1.61% | Spanish |
| 18 | 164 | 1.56% | English |
| 19 | 161 | 1.53% | English |
| 20 | 149 | 1.42% | Spanish |
| 22 | 147 | 1.40% | English |
| 21 | 147 | 1.40% | Spanish |
| 23 | 129 | 1.23% | English |
| 24 | 122 | 1.16% | English |
| 25 | 105 | 1.00% | English |

Source: Author's own work.

First of all, in the network of TikTok comments, the positions are not so clearly defined as in the network of Twitter retweets (Table 3). In the analysis of the most important communities on this platform, conducted by means of the number and the percentage of nodes they contain, the implicit position in their content and the main content creators indicate that there continues to be a degree of debate in communities on either side of the dividing line.

This dialogue is not always of great quality, since there is considerable caricaturing and even ad hominem arguments, but nonetheless, unlike on Twitter, the exchange of arguments for and against the position of each group is more present on TikTok.

On TikTok, those communities supporting trans-inclusive feminism, unlike the more aggressive approach in the Twitter discourse, denounce the violence to which they are subjected at the hands of non-trans-inclusive activists. There is also plenty of the content typical of TikTok, caricaturing the adversary, or even videos in response to other videos posted by proponents of the opposing position. It should nonetheless be emphasized that although the opinions attributed to the interlocutor are often distorted, ridiculed and simplified by both communities, TikTok and its comments much more closely resemble the supposed town square for debate that so many authors hoped Twitter would serve as more than a decade ago now. This is also reflected in the existence of communities in which multiple discourses coexist, which we have referred to as mixed communities. In terms of language, there is a substantial presence of Spanish as the language of the "TERF" debate, indicating the increasing prominence of the debate in Spain and Latin American countries.

## 4. Conclusions and discussion

The debate centred on the use of the term "TERF" illustrates firstly the considerable polarization caused within feminist activism on social media by the inclusion within this context of the claims of the trans movement. The debate as to the political subject of feminist claims is a complex and long-standing one, which has been reinforced by its transfer to digital spaces (Earles, 2017). The study of Twitter and TikTok served to identify communities with considerable cohesion, little porosity, characterized by a high degree of homophily, and with very little dialogue between them. Polarized and aggressive, even dialectically violent, conversations can likewise be found, with a substantial degree of confrontation (Williams, 2020; Ferré-Pavia & Zaldívar, 2022).

Secondly, the so-called "TERF wars" on social media are, to a great extent, an expression of the intergenerational conflict which exists within feminism, and which is linked to disputes between members of the second and third waves, as opposed to the fourth wave of the movement (Maulding, 2019). References to age may also be found running through some of the references made. By age group, younger users on Twitter and TikTok opt, to a greater extent, for trans-inclusive positions, as opposed to older groups, who maintain a largely opposing position. Interactive elements allow the hyper-personalization of the messages and reinforce their meaning, an opportunity exploited by young people as a group, who represent the majority on such platforms, to exert their own space of influence in the debate and discussion of topics with which they feel social and affective engagement. We should again here consider the presence of different age groups on social media, and the way in which this influences the perception of these debates (Schuster, 2013).

Thirdly, as a result of the majority participation of young women in the debates on these social media platforms, the dominant positions on Twitter and TikTok are aligned with trans-inclusive feminist activism. Meanwhile, support for these positions is included more horizontally across content associated with other types of topic, unlike in non-inclusive communities, which are the minority, and have a more focal and specific nature.

It was lastly possible to confirm differential use between the two social media platforms, with TikTok being a less partisan and more dialogue-based network than Twitter, which lends itself to discussions and participations in a more neutral tone. This may be influenced by factors such as a younger user profile (which means that an intergenerational debate is less present on the platform), the focus on entertainment in most of the content (Peña-Fernández et al., 2022) and the abundance of non-textual elements. On Twitter, meanwhile, there is a greater presence of strategies and dynamics of polarized discourse and segregation, perhaps as a result of the use of algorithms which give rise to the formation of echo chambers and closed communities (Cho et al., 2020). This effect could generate a greater degree of polarization in areas that necessarily require integrating and inclusive perspectives, as is the case of issues connected with gender, feminism and the LGTBI+ movement, and be harmful for dialogue, deliberation and understanding between different positions as to the subject and object of feminism. This influence had to date been linked mainly to debates and topics in the sphere of politics, rather than issues connected with gender and sexual identity, nor had it focused on a younger population group. In line with Banks et al. (2018), it is also possible to deduce that repeated exposure to trans-inclusive messages online, above all on Twitter, as well as signs of confrontation itself, would be helping to reinforce attitudinal positions in the opposing direction (Woolley & Howard, 2018).